\documentclass[preprint,showpacs,superscriptaddress,aps]{revtex4}
\usepackage{amssymb}
\usepackage{amsmath,amssymb,graphicx}
\usepackage{graphicx}
\usepackage{dcolumn}
\usepackage{bm}
\def\be{\begin{equation}}
\def\ee{\end{equation}}
\def\bea{\begin{eqnarray}}
\def\eea{\end{eqnarray}}

\def\lsim{\raise0.3ex\hbox{$\;<$\kern-0.75em\raise-1.1ex\hbox{$\sim\;$}}}
\def\gsim{\raise0.3ex\hbox{$\;>$\kern-0.75em\raise-1.1ex\hbox{$\sim\;$}}}

\usepackage{pstricks}
\begin{document}
\title{{\fontsize{13}{13}\selectfont{TeV Scale Leptogenesis in $B-L$ Model with Alternative Cosmologies}}}
\author{W. Abdallah}
\email{wabdallah@zewailcity.edu.eg}
\affiliation{{\fontsize{10}{10}\selectfont{Centre for Theoretical Physics, Zewail City of Science and Technology, Sheikh Zayed, 12588, Giza, Egypt.}}}
\affiliation{{\fontsize{10}{10}\selectfont{Department of Mathematics,  Faculty of Science, Cairo University, Giza, Egypt.}}}
\author{D. Delepine}
\email{delepine@fisica.ugto.mx}
\affiliation{{\fontsize{10}{10}\selectfont{Division de Ciencias e Ingenier\'ias,  Universidad de Guanajuato, C.P. 37150, Le\'on, Guanajuato, M\'exico.}}}
\author{S. Khalil}
\email{skhalil@zewailcity.edu.eg}
\affiliation{{\fontsize{10}{10}\selectfont{Centre for Theoretical Physics, Zewail City of Science and Technology, Sheikh Zayed, 12588, Giza, Egypt.}}}
\affiliation{{\fontsize{10}{10}\selectfont{Department of Mathematics, Faculty of Science, Ain Shams University, Cairo, Egypt.}}}
\begin{center}
\begin{abstract}
In TeV scale $B-L$ extension of the standard model with inverse
seesaw, the Yukawa coupling of right-handed neutrinos can be of
order one. This implies that the out of equilibrium condition for leptogenesis within
 standard cosmology is not satisfied.  We provide two scenarios
for overcoming this problem and generating the desired value of
the baryon asymmetry of the Universe. The first scenario is based
on extra-dimensional braneworld effects that modify the Friedman
equation. We show that in this case the value of the baryon
asymmetry  of the Universe constrains the five-dimensional Planck
mass to be of order ${\cal O}(100)$ TeV. In the second scenario a
non-thermal right-handed neutrino produced by the decay of
inflaton is assumed. We emphasize that in this case, it is
possible to generate the required baryon asymmetry of the Universe
for TeV scale right-handed neutrinos.
\end{abstract}
\end{center}
\maketitle
\section{Introduction}

The Standard Model (SM) of electroweak and strong interactions has
had a tremendous success when confronted with experiments. However,
non-vanishing neutrino masses provides the first confirmed hint
towards physics beyond the SM. The evidence of very light neutrino
masses is now well established by measuring neutrino oscillations
in solar and atmospheric neutrinos. It has been  shown that the
minimal extension of the SM gauge group by an extra $U(1)$ $B-L$
gauge symmetry has all the necessary requirements to generate the
observed neutrino masses even if the right-handed neutrino masses
are  around TeV scale
\cite{Mohapatra:1980qe,Marshak:1979fm,Wetterich:1981bx,Masiero:1982fi,Mohapatra:1982xz,Buchmuller:1991ce,Abbas}.
In particular, this type of models has three SM singlet fermions
that arise as a result of the anomaly cancellation conditions.
These particles account for right-handed neutrinos and give a
natural explanation for the seesaw mechanism. This simple
extension of the SM predict an extra SM singlet scalar and an
extra  neutral gauge boson associated to $B-L$ gauge symmetry.
These new particles may have significant impact on the SM
phenomenology, leading to interesting signatures at Large
Hadron Collider (LHC)
\cite{Khalil:2006yi,Emam:2007dy,Huitu:2008gf,Blanchet:2009bu,Basso:2011hn}.

On the other hand, the observed Baryon Asymmetry in the Universe (BAU)
provides another indication for physics beyond the SM since it has
been well  established that the strength of the CP violation in
the SM is not sufficient to generate this asymmetry. The CP
violating decay of the right handed neutrinos may generate a
lepton asymmetry which is transformed through the SM sphalerons
processes to a baryon asymmetry \cite{Fukugita:1986hr,Covi:1996wh,buchmuller,Buchmuller:2004nz}. This scenario is called
leptogenesis which is very attractive at light of the neutrino
mixings parameters precision measurements.

From Big Bang nucleosynthesis, the bound on the BAU, defined as $\eta_B \equiv n_B/n_{\gamma}$
where $n_{B,\gamma}$ are respectively the baryon  and photon
density, is given by \cite{Steigman:2010zz}
\begin{equation}
\eta_B=6.07 \pm 0.33\times 10^{-10},
\end{equation}
which is consistent with the result reported recently by
WMAP7 \cite{Larson:2010gs}
\begin{equation}
\eta_B=6.160^{+0.153}_{-0.156}\times 10^{-10}.
\end{equation}
In Ref. \cite{Abbas}, it was shown that within the $B-L$ model
with type I seesaw, where the Yukawa coupling of right-handed
neutrinos $\lambda_\nu$ is of order $ {\cal O}(10^{-6})$, the
lepton asymmetry is a few order of magnitude below the recent
observed value of the BAU. The assumption of strong mass
degeneracy between the first two right-handed neutrinos was
considered as a possible approach for enhancing this asymmetry. It
turns out that the mass difference between right-handed neutrinos,
$\triangle M$, must be of order ${\cal O}(10^{-3})\;\text{GeV}$ in
order to have a successful leptogenesis. Such small difference
masses may be considered as unnatural fine-tuning.

The $B-L$ model with Inverse Seesaw (ISS), where neutrino Yukawa
coupling is of order one, has been also analyzed
\cite{Khalil:2010iu}. In this class of models, the large coupling
between heavy neutrinos and SM particles leads to interesting
implications \cite{Abdallah:2011M} and enhance the accessibility
of TeV scale $B -L$ at the LHC. However, in the Standard
Cosmology(SC), one can easily show that within TeV scale $B-L$
with ISS, it is not possible to produce the BAU through thermal
leptogenesis. The reason is that the out-of-equilibrium condition
\cite{Kolb} for the decay of the right-handed neutrinos, which
prevents the generated asymmetry from being washed out by the
inverse decays and scattering processes, is not satisfied.

In this paper, we show that it is possible to implement a
successful leptogenesis in these $B-L$ models with ISS without
fine-tuning. We consider two examples to avoid the exponentially
suppression of the lepton asymmetry in these models. One is to
change the Universe Dynamics. For instance, assuming our world to
be trapped on a brane. In this non-standard cosmology, the
extra-dimensional braneworld  effects modify the Friedmann
equation that governs the cosmological evolution of our Universe.
The main result is to modify the expansion rate of the Universe.
So, using the baryon asymmetry of the Universe measurement, it is
possible to constraint the five dimensional Planck Mass scale
$M_5$. The second solution to our problem is to produce the lepton
asymmetry through non-thermal leptogenesis. In this case, we show
how this scenario can be easily implemented in these $B-L$ models
and that it is possible to produce the right order of magnitude
for the BAU with an inflation mass scale around
$10^{12}\;\text{GeV}$ even with right-handed neutrino masses
around TeV scale.

This paper is organized as follows: In section 2, the main
characteristics of the $B-L$ model with ISS and their problems
with thermal leptogenesis are reviewed.  Leptogenesis in a
Braneworld Cosmology (BC) is studied in section 3 where we show
that to produce the right order of magnitude of the BAU, the $M_5$
scale should be of order $10^{5}\; \text{GeV}$. In section 4, the
non-thermal leptogenesis scenario is implemented in this type of
$B-L$ models. Finally our conclusions are given in section 5.

\section{{\fontsize{10}{10}\selectfont{$B-L$ with Inverse seesaw mechanism and thermal leptogenesis}}}

As advocated in the introduction, non-vanishing neutrino masses
represent a firm observational evidence of new physics beyond the
SM. The $B-L$ extension of the SM permit to introduce naturally
three right-handed neutrinos due to anomaly cancelation condition
and therefore to explain the light neutrino masses through see-saw
mechanism. The model is based on the gauge group $SU(3)_C\times
SU(2)_L\times U(1)_Y\times U(1)_{B-L}$. The Lagrangian of the
leptonic sector is given by \cite{Khalil:2010iu}
\begin{eqnarray}
{\cal L}_{B-L}&=&-\frac{1}{4} F'_{\mu\nu}F'^{\mu\nu} + i~ \bar{\ell}_L
D_{\mu} \gamma^{\mu} \ell_L + i~ \bar{e}_R D_{\mu} \gamma^{\mu} e_R + i~
\bar{N}_R D_{\mu} \gamma^{\mu} N_R\nonumber\\
&+& i~ \bar{S}_{1} D_{\mu} \gamma^{\mu} S_{1} + i~ \bar{S}_{2}
D_{\mu} \gamma^{\mu} S_{2}
+(D^{\mu}\phi)^\dagger D_{\mu} \phi + (D^{\mu} \chi)^\dagger D_{\mu}\chi -V(\phi, \chi)\nonumber\\
&-&\Big(\lambda_e \bar{\ell}_L \phi e_R+\lambda_{\nu} \bar{\ell}_L
\tilde{\phi} N_R +\lambda_{N} \bar{N}^c_R \chi S_2+ h.c.\Big) -
\frac{1}{M^3}\bar{S}^c_{1} {\chi^\dag}^{4}
S_{1}-\frac{1}{M^3}\bar{S}^c_{2} {\chi}^{4} S_{2},
\end{eqnarray}
where $F'_{\mu\nu}$ is the field strength of the $U(1)_{B-L}$,
$D_{\mu}$ are the respective covariant derivatives,
$N^i_R,\;i=1,2,3$; are the right-handed neutrinos with $B-L$
charge $=-1$, and $S^{i}_{j},\;i=1,2,3,\;j=1,2$ are two SM singlet
fermions, each of them has three flavors and their $B-L$ charges
$=\mp2$. In addition, $\chi$ is an extra SM singlet scalar with
$B-L$ charge equal to one and $\phi$ is the usual electroweak
Higgs fields. In general, the scale of $B-L$ symmetry breaking is
unknown, ranging from TeV to much higher scales. However, it was
proven that in supersymmetric framework, the scale of $B-L$ is
nicely correlated with the soft supersymmetry breaking scale,
which is TeV \cite{Khalil}. Therefore, to be consistent with the
result of radiative $B-L$ symmetry breaking found in gauged $B-L$
model with supersymmetry, we assume that the non-vanishing vacuum
expectation value (VEV) of $\chi:|\langle\chi\rangle|=
v'/\sqrt{2}$ is to be of order TeV. After $B-L$ symmetry breaking,
the mass terms for the right-handed neutrinos are given by
\begin{eqnarray}
{\cal L}_m^{\nu}=\mu_s \bar{S}^c_{2} S_{2}+\left(M_N \bar{N}^c_R S_2+h.c.\right),
\end{eqnarray}
where $$M_N=\frac{1}{\sqrt 2}\lambda_N\;
v',\;\;\;\;\;\;\;\mu_s=\frac{v'^4}{4 M^3}\sim10^{-10}\; \text{GeV}.$$
Therefore, in the basis $\{N_R, S_{2}\}$, the $6\times 6$
neutrinos mass matrix takes the form
$${\cal M}_{\nu}=
\left(%
\begin{array}{c|cc}
& N_R & S_{2}\\\hline
\bar{N}^c_R& 0 & M_N \\[0.2cm]
\bar{S}^c_{2}& M^T_N & \mu_s\\
\end{array}%
\right).$$ %
Thus, by diagonalizing ${\cal M}_{\nu}$ one can obtain
the following heavy neutrinos masses
\begin{eqnarray}
m_{\nu_{H,H'}}=\frac{1}{2} \left(\mu_s \mp\sqrt{\mu_s^2+4 M_{N}^2}\right)
\end{eqnarray}
corresponding to  the following  mass eigenstates
\begin{equation}
\nu_{H,H'}\simeq\mp\frac{1}{\sqrt{2}}N+\frac{1}{\sqrt{2}}S_2.
\end{equation}
Eventually, heavy neutrinos will decay, and the total lepton asymmetry is generated due to the CP asymmetry that arises through the interference
of the tree level and one-loop diagrams as usual in leptogenesis scenario \cite{Fukugita:1986hr,buchmuler,Covi:1996wh}
\begin{eqnarray}
\epsilon=\epsilon_1+\epsilon_{1'}=\frac{\sum_\alpha{\left[\Gamma(\nu_{H_1}\rightarrow
\phi^*\,\ell_\alpha)-\Gamma(\nu_{H_1}\rightarrow
\phi\,\bar{\ell}_\alpha)\right]}}{\sum_\alpha{\left[\Gamma(\nu_{H_1}\rightarrow
\phi^*\,\ell_\alpha)+\Gamma(\nu_{H_1}\rightarrow
\phi\,\bar{\ell}_\alpha)\right]}} +(\nu_{H_1} \rightarrow
\nu_{H'_1}),
\end{eqnarray}
where $\epsilon_1, \epsilon_{1'}$ are the lepton asymmetries due
to the decay of the lightest heavy neutrinos $\nu_{H_1}$ and
$\nu_{H'_1}$, respectively. For simplicity and due to
$|m_{\nu_{H_i}}|\simeq |m_{\nu_{H'_i}}|$, we shall assume that
both contributions are of the same order of magnitude ($\epsilon_1
\simeq \epsilon_1^{'}$). Therefore, one  obtains
\begin{equation}\label{eq3}
\epsilon\simeq\frac{1}{4\pi}\frac{1}{(\lambda_\nu\lambda^\dag_\nu)_{11}}\sum_{k=2,3}{\text{Im}
\left[(\lambda_\nu\lambda^\dag_\nu)^2_{1k}\right]\left[f\left(\frac{m^2_{\nu_{H_k}}}{m^2_{\nu_{H_1}}}\right)+g\left(\frac{m^2_{\nu_{H_k}}}{m^2_{\nu_{H_1}}}\right)\right]},
\end{equation}
where
\begin{eqnarray}
f(x_k)&=& \sqrt{x_k}\left[1-(1+x_k)\ln\left(\frac{1+x_k}{x_k}\right)\right],\\
g(x_k)&=& \frac{\sqrt{x_k}\,(1-x_k)}{\large\vert1-x_k-a(m^2_{\nu_{H_1}})[(\lambda_\nu^\dagger\lambda_\nu)_{kk}-\sqrt{x_k}(\lambda_\nu^\dagger\lambda_\nu)_{11}]\large\vert^2},\\
a(q^2)&=&\frac{1}{16 \pi^2}\left(\ln \frac{q^2}{\mu^2}-2-i\pi \Theta(q^2)\right),
\end{eqnarray}
with $x_k=\frac{m^2_{\nu_{H_k}}}{m^2_{\nu_{H_1}}}$. For $\vert m_{\nu_{H_k}}-m_{\nu_{H_1}}\vert\gg\vert\Gamma_{D_i}-\Gamma_{D_1}\vert$, $g(x_k)$ is given by
\begin{equation} %
g(x_k)= \frac{\sqrt{x_k}}{1-x_k}.%
\end{equation}
Therefore, the final $CP$ asymmetry  is given by
\begin{eqnarray}\label{eq7}
\epsilon&\simeq& \frac{1}{4\pi}\frac{1}{(\lambda_\nu\lambda^\dag_\nu)_{11}}\sum_{k=2,3}{\text{Im}
\left[(\lambda_\nu\lambda^\dag_\nu)^2_{1k}\right]\sqrt{x_k}\left[1+\frac{1}{1-x_k}-(1+x_k)\ln\left(\frac{1+x_k}{x_k}\right)\right]}.
\end{eqnarray}
Within ISS mechanism, the Dirac neutrino Yukawa couplings,
$\lambda_\nu$, can be written as \cite{Abdallah:2011M}
\be %
\lambda_\nu=\frac{1}{v}\; U_{\!M\!N\!S}\, \sqrt{m_{\nu_l}^{\rm diag}}\, R\, \sqrt{\mu^{-1}_s}\, M_N, %
\label{mD}
\ee %
where $v=174\; \text{GeV}$ is the VEV of the Higgs field $\phi$,
$U_{\!M\!N\!S}$ is the physical neutrino mixing matrix \cite{mns},
$m_{\nu_l}^{\rm diag}$ is the diagonal matrix of the light
neutrino masses and $R$ is an arbitrary orthogonal matrix. For
example, if $\mu_{s_{1}}=2.2\times10^{-12}\;{\rm GeV},\,
\mu_{s_{2}}=4.1\times10^{-10}\;{\rm
GeV},\,\mu_{s_3}=5\times10^{-8}\;{\rm GeV},\, m_{\nu_{H_1}}=1\,{\rm
TeV},\, m_{\nu_{H_2}}=1.5\,{\rm TeV},\, m_{\nu_{H_3}}=2\,{\rm TeV},\,
m_{\nu_{l_{1}}}=10^{-13}\; {\rm GeV},\,
m_{\nu_{l_{2}}}=8.74\times10^{-12}\;{\rm GeV},\,
m_{\nu_{l_{3}}}=4.95\times10^{-11}\;{\rm GeV}$, then one finds
that $\epsilon\simeq -1\times10^{-4}$. For prevents the generated
asymmetry given in (\ref{eq7}) from being washed out by the
inverse decay and scattering processes mediated by $\nu_{H_1}$, it
is useful to define a quantity
\begin{eqnarray}
K=\frac{\Gamma_{D_1}}{H}{\Big{|}}_{T=m_{\nu_{H_1}}},
\end{eqnarray}
where $H$ is the expansion rate of the Universe and $\Gamma_{D_1}$
is the total decay rate of $\nu_{H_1}$. The wash out effects due
to the inverse decay and the scattering processes is parameterized
by a coefficient $\eta$ which depends on $K$ parameter, and the
final amount of lepton asymmetry is given by
\begin{eqnarray}
Y_L\equiv\frac{n_L-n_{\bar{L}}}{s}=\eta\frac{\epsilon}{g_{*}},
\end{eqnarray}
where $s$ is the entropy density, $g_{*}$ is the number of relativistic degrees of freedom and $g_{*}\simeq 106.75$ for the SM and $n_L$ is the lepton number density.
Finally, the electroweak sphalerons  convert the lepton asymmetry $Y_L$ to baryon asymmetry $Y_B$ through the usual conversion factor $c$:
\begin{eqnarray}
Y_B=\frac{c}{c-1}Y_L\simeq-1.4\times10^{-3}\;\eta\;\epsilon.
\end{eqnarray}
The amount of wash out parameterized by $\eta$ depends on the size of $K$ as the following \cite{Kolb,Buchmuller:2004nz}:
\begin{enumerate}
\item If $K \lesssim{\cal O}(1)$, the inverse decay and the scattering processes are important and $\eta\thicksim1$.
This is known as out-of-equilibrium condition. Thus the net $Y_B$ is
    \begin{eqnarray}\label{eq4}
    Y_B\simeq-1.4\times10^{-3}\;\epsilon.
    \end{eqnarray}
\item If ${\cal O}(1) \lesssim K\lesssim 10^6$, then during the
epoch when $B$-nonconserving processes are effective, the inverse
decays are more important in damping the baryon asymmetry than the
scattering processes. Thus, for this range of $K$ the parameter
$\eta$ is given by $\eta\simeq\frac{0.3}{K(\ln K)^{0.6}}$, hence
the net $Y_B$ is
    \begin{eqnarray}\label{eq5}
    Y_B\simeq-1.4\times10^{-3}\;\;\frac{0.3\;\epsilon}{K(\ln K)^{0.6}}.
    \end{eqnarray}
\item When $K > 10^6$, the scattering processes are the dominant $B$ damping processes and there is no departure from thermal equilibrium and as a consequence the net lepton asymmetry vanishes because $\eta$ decrease exponentially (with $K^{1/4}$). Thus, the final baryon asymmetry does decrease exponentially.
\end{enumerate}
As known in the SC, the expansion rate of the Universe $H$ is given by
\begin{equation}
H\simeq1.66\sqrt{g_{*}}\frac{T^2}{M_{pl}}.
\end{equation}
Therefore, according to our model$\;(m_{\nu_{H_1}}\sim{\cal O}(\text{TeV}),\;\lambda_\nu\sim{\cal O}(1)),$ $H$ is of order $10^{-12}$, thus $K>10^6,$ and the
final baryon asymmetry does decrease exponentially as mentioned
above. In the next section, we consider possible scenario to overcome this problem.
\section{Leptogenesis with Braneworld Cosmology}
Standard leptogenesis scenario in non-standard cosmology has
already been studied in Ref. \cite{Bento:2005je,Okada:2005kv}.
In these cosmological models, the $H$ parameter is
changed and it is expected that these modifications could make
more efficient leptogenesis process dumping the wash-out
processes. The extra-dimensional braneworld effects modify the
Friedmann equation that governs the cosmological evolution of a
FRW Universe trapped on the brane. For instance, in case of
Randall-Sundrum type II braneworld model \cite{Randall:1999vf},
the modified Friedmann equation is
\begin{eqnarray}\label{eq2}
H^2=\frac{8\pi}{3
M^2_{Pl}}\rho\left(1+\frac{\rho}{2\lambda}\right),\;\;\;\;\;\;
\lambda=\frac{3}{4\pi}\frac{M^6_5}{M^2_{Pl}},
\end{eqnarray}
where $M_{Pl}\simeq1.22\times10^{19}$ GeV is the four-dimensional
Planck mass, $M_5$ is the five-dimensional Planck mass, $\rho$ is
the energy density of the matter degrees of freedom trapped in the
brane, and we have set the four-dimensional cosmological constant
to zero and assumed that inflation rapidly makes any dark
radiation term negligible as it is strongly constrained by
nucleosynthesis \cite{Ichiki:2002eh}. For a radiation era one can
write $\rho$ in terms of the temperature:
\begin{eqnarray*}
\rho=\frac{\pi^2}{30}g_{*}T^4.
\end{eqnarray*}
From these equations, it is possible to define a "transition
temperature" $T_t$ which is define as $\rho(T_t)/2\lambda=1$, at
which the evolution of the early Universe changes from the BC era into the standard one. The transition
temperature is given by
\begin{equation}
T_t\simeq 1.6 \times 10^{7}\left(\frac{100}{g_{*}}\right)^{1/4}\left(\frac{M_5}{10^{11}\; \text{GeV}}\right)^{3/2} \text{GeV}.
\end{equation}
The constraints on $\lambda$ parameter essentially come from Big
Bang nucleosynthesis which implies that $\lambda >(1\;\text{MeV})^4$ which
corresponds to $M_5 > 8.8\; \text{TeV}$ \cite{Langlois:2002bb}.
\begin{figure}[t]
\begin{center}
\includegraphics[width=11cm,height=7cm]{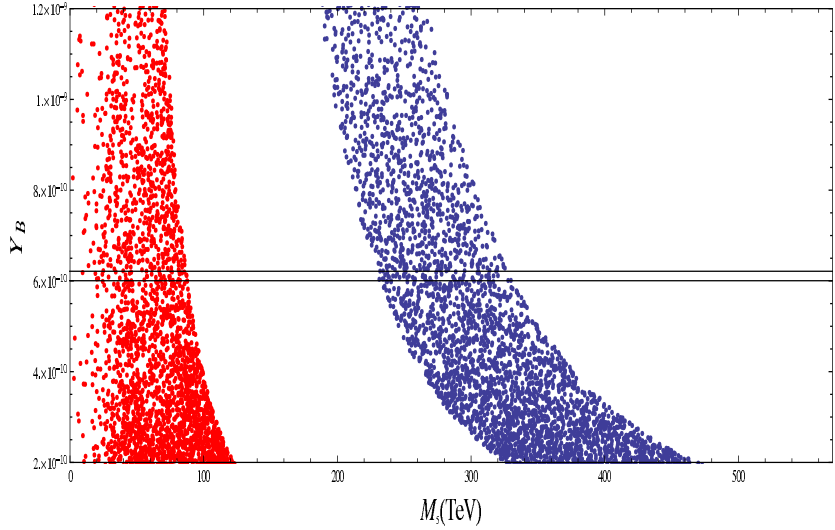}
\caption{Baryon asymmetry in $B-L$ extension of the SM with ISS versus the five-dimensional Planck mass $M_5$, where $m_{\nu_{H_1}}\in[1,1.1]\,\text{TeV}$ for the left points (Red) and $m_{\nu_{H_1}}\in[1.3,1.4]\,\text{TeV}$ for the right ones (Blue). The
horizontal lines refer to the WMAP7 result limit of $\eta_B$. The other parameters are fixed as
follows: $\mu_{s_{1}}=2.2\times10^{-12}\;{\rm GeV},\, \mu_{s_{2}}=4.1\times10^{-10}\;{\rm GeV},\,\mu_{s_3}=5\times10^{-8}\;{\rm GeV},\, m_{\nu_{H_2}}=1.5\,{\rm TeV},\, m_{\nu_{H_3}}=2\,{\rm TeV},\,
m_{\nu_{l_{1}}}=10^{-13}\; {\rm GeV},\,m_{\nu_{l_{2}}}=8.74\times10^{-12}\;{\rm GeV},\,
m_{\nu_{l_{3}}}=4.95\times10^{-11}\;{\rm GeV}$. }
\label{Y1}
\end{center}
\end{figure}
Notice that the modified Friedmann equation (\ref{eq2}) reduces to
the usual Friedmann equation (as in the case of the SC) at
sufficiently low energies, $\rho\ll2\lambda \Rightarrow
H\propto\sqrt{\rho}$, while at very high energies we have
$\rho\gg2\lambda \Rightarrow H\propto\rho$, as in the case of the BC. Now, for the BC, the expansion rate of
the Universe $H$ is given by
\begin{equation}
H\simeq1.38\,g_{*}\frac{T^4}{M^3_5}.
\end{equation}
Therefore, $H$ can be enhanced from $10^{-12}$ to $10^{15}
M_5^{-3}$, i.e. for $M_5 \simeq {\cal O}(10)$ TeV the ratio $K$
can be of order ${\cal O}(10)$, hence the required baryon
asymmetry $Y_B\simeq {\cal O}(10^{-10})$ can be obtained with
lepton asymmetry of order $10^{-3}$, according to Eq.(\ref{eq5}).
It is interesting to note that using the observed value for the
BAU is possible to constraint the $M_5$ scale. As shown in Fig.
\ref{Y1}, $M_5$ has to be of order ${\cal O}(100)$ TeV to be
within the allowed region for the BAU.  These results have to be
compared to standard leptogenesis scenario in the BC
\cite{Bento:2005je,Okada:2005kv} where typically the $M_5$ scale
has to be as large as $10^{10}$ GeV. For our values of $M_5$, one
expects a low value for the transition temperature (around 10-100
MeV). With these relatively low value of $M_5$ scale, one can
expect strong effects from the BC on possible observables as dark
matter relic density (see for instance Ref. \cite{Nihei:2004xv}).
\section{TeV Non-Thermal Leptogenesis}

Another way to overcome the problem of thermal leptogenesis in
these kind of models is to look for a non-thermal production of
the heavy right handed neutrinos. In such scenario, the out of
equilibrium condition is generate through reheating once the
inflation decays \cite{Lazarides:1991wu,Asaka:1999yd,Asaka:1999jb,Giudice:1999fb, Pallis:2012iw}. To be possible to have a non-thermal production of the lightest right-handed (RH) neutrino, we assume that an inflaton decays
dominantly into a pair of lightest RH neutrinos,
$\chi'\rightarrow\nu_{H_1}\,\nu_{H_1}$. For this decay to occur,
the inflaton mass $M_{\chi'}$ has to be greater than
$2\,m_{\nu_{H_1}}$. The inflaton field $\chi'$ is a scaler field
singlet under the SM gauge group with $B-L$ charge=0, and it has
the following gauge invariant terms
\[{\cal L}_{\chi'}=\dots-\lambda_{\chi'} \bar{N}_R\chi'N_R+m^2\chi'^\dagger\chi',\;\;\;
\text{where}\;\;\; m^2>0.\]
The reheating temperature $T_R$ following the inflationary epoch is given by
\be %
T_R=\left(\frac{90}{8\pi^3 g_*}\right)^\frac{1}{4}\left(\Gamma_{\chi'}M_{pl}\right)^{\frac{1}{2}},
\ee
where \[\Gamma_{\chi'}=\frac{1}{4\pi}|\lambda_{\chi'}|^2 m_{\nu_{H_1}}\]
is the decay width of the inflaton field $\chi'$ according to its decay $\chi'\rightarrow\nu_{H_1}\,\nu_{H_1}$.
For inflaton mass is of order $10^{12} \;\text{GeV}$, one finds that $T_R \simeq 10^9\;\text{GeV}$. \\[0.3cm]
\begin{figure}[t]
\begin{center}
\includegraphics[width=12cm,height=8cm]{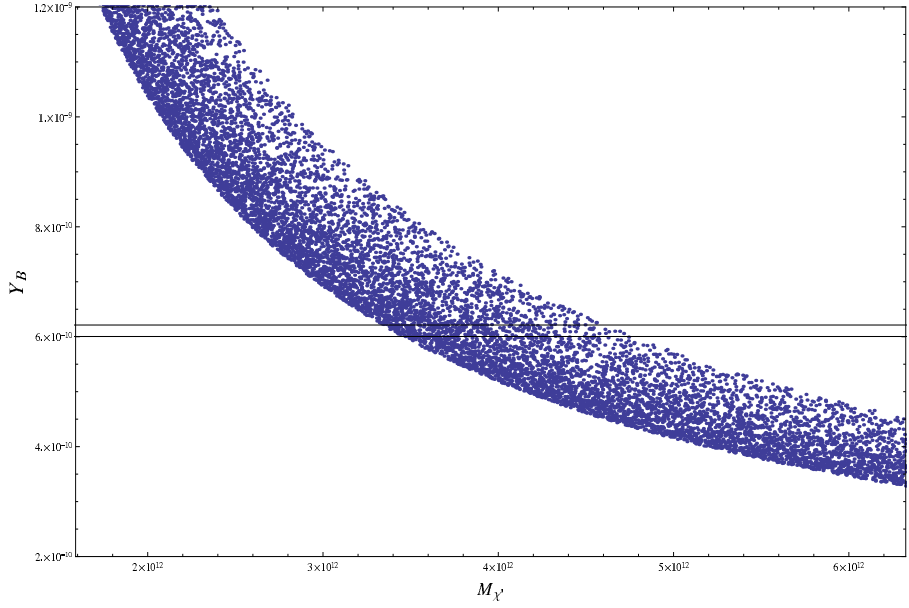}
\caption{Baryon asymmetry as function of the inflaton mass $M_{\chi'}$
within the non-thermal leptogenesis, for $m_{\nu_{H_1}} = 1\,\text{TeV}$, $m_{\nu_{H_2}}= 1.5\,\text{TeV}$,
$m_{\nu_{H_3}}\in [1.5,2.5]\,\text{TeV}$. The horizontal lines refer to the WMAP7 result limit of $\eta_B$. The other parameters are fixed as in Fig. \ref{Y1}.}
\label{Y2}
\end{center}
\end{figure}
The  decays $\chi'\rightarrow \nu_{H_1}\,\nu_{H_1} $ reheats the Universe and the subsequent decays of the
right-handed neutrino produced in this way generate a Lepton asymmetry given by
\be %
Y_L = \frac{3}{2} BR(\chi' \to \nu_{H_1}\,\nu_{H_1})
\frac{T_R}{M_{\chi'}} \varepsilon,\nonumber %
\ee %
where $\epsilon$ is the CP asymmetry given in Eq. (\ref{eq7}).
Since $BR(\chi' \to \nu_{H_1}\,\nu_{H_1})\simeq {\cal O}(1)$, and
$T_R/M_{\chi'}\simeq {\cal O}(10^{-3})$, the non-thermal lepton
asymmetry can be suppressed with about three order of magnitudes,
respect to the thermal lepton asymmetry. Thus the baryon asymmetry
$Y_B$ in the non-thermal scenario is of order $10^{-10}$, as
required. In Fig. \ref{Y2}, we present the baryon asymmetry in
non-thermal scenario as function of the inflaton mass. Here we
assume that the right-handed neutrino masses are given by
$m_{\nu_{H_1}}=1$ TeV, $m_{\nu_{H_2}}=1.5$ TeV, and
$m_{\nu_{H_3}}\in[1.5,2.5]$ TeV. The Dirac neutrino Yukawa
couplings, $\lambda_\nu$, are defined as in Eq. (\ref{mD}). It is
interesting to note that for $M_{\chi^{'}}$ of order $10^{12}$ GeV
is possible to explain naturally the BAU in these $B-L$ models
with ISS.
\section{Conclusion}

In TeV scale $B-L$ extension of the standard model with ISS, the
Yukawa coupling of right-handed neutrinos interactions
$\lambda_\nu\sim 1$ and it implies that  in standard cosmology,
the out of equilibrium condition for the inverse decay of the
right-handed neutrinos and $\Delta L=2$ scattering processes,
which prevents the generated lepton asymmetry from being washed
out is not satisfied. In this paper, we have studied two ways to
solve this problem and to generate the desired baryon asymmetry of
the Universe. Firstly, we considered that extra-dimensional
braneworld effects modify the Friedman equation that governs the
cosmological evolution of the FRW Universe trapped on the brane,
it is possible to strongly constraint the five-dimensional Planck
mass $M_5$ to be of order ${\cal O}(100)$ TeV in order to produce
the right amount of lepton asymmetry needed to generate the BAU.
It is important to emphasize that a such approach can be
generalized to any alternative cosmology and the precise
measurement of the BAU can be used to constraint these alternative
to standard cosmology. Secondly, we have shown that in a
leptogenesis scenario based on non-thermal right-handed neutrinos
decays  it is possible to generate the matter-antimatter asymmetry
of the Universe with an inflaton mass of order $10^{12}$ GeV even
if the right-handed majorana neutrino scale is around 1 TeV.
\section*{Acknowledgements}
The work of W. A. and S. K. is partially supported by the ICTP grant
AC-80. D. D. is  grateful to Conacyt (M\'exico), DAIP project
(Guanajuato University) and PIFI (Secretaria de Educacion Publica,
M\'exico) for financial support.

\end{document}